# The Effect of Temperature on Cu-K-In-Se Thin Films

Christopher P. Muzzillo,[1,2] Ho Ming Tong,[2,3] and Timothy J. Anderson[2]

[1]National Renewable Energy Laboratory, Golden, CO 80401, USA

[2]Department of Chemical Engineering, University of Florida, Gainesville, FL 32611, USA

[3]Oak Ridge National Laboratory, Oak Ridge, TN 37831, USA

Abstract

Films of Cu-K-In-Se were co-evaporated at varied K/(K+Cu) compositions and substrate temperatures (with constant (K+Cu)/In ~ 0.85). Increased Na composition on the substrate's surface and decreased growth temperature were both found to favor $Cu_{1-x}K_xInSe_2$ (CKIS) alloy formation, relative to mixed-phase $CuInSe_2$ + $KInSe_2$ formation. Structures from X-ray diffraction (XRD), band gaps, resistivities, minority carrier lifetimes and carrier concentrations from time-resolved photoluminescence were in agreement with previous reports, where low K/(K+Cu) composition films exhibited properties promising for photovoltaic (PV) absorbers. Films grown at 400-500°C were then annealed to 600°C under Se, which caused K loss by evaporation in proportion to initial K/(K+Cu) composition. Similar to growth temperature, annealing drove CKIS alloy consumption and $CuInSe_2$ + $KInSe_2$ production, as evidenced by high temperature XRD. Annealing also decomposed $KInSe_2$ and formed $K_2In_{12}Se_{19}$. At high temperature the $KInSe_2$ crystal lattice gradually contracted as temperature and time increased, as well as just time. Evaporative loss of K during annealing could accompany the generation of vacancies on K lattice sites, and may explain the $KInSe_2$ lattice contraction. This knowledge of Cu-K-In-Se material chemistry may be used to predict and control minor phase impurities in Cu(In,Ga)(Se,S)$_2$ PV absorbers—where impurities below typical detection limits may have played a role in recent world record PV efficiencies that utilized KF post-deposition treatments.

Introduction

Recent reports have detailed power conversion efficiency enhancements when potassium fluoride and selenium have been co-evaporated onto Cu(In,Ga)(Se,S)$_2$ (CIGS) absorbers at around 350°C (KF post-deposition treatment (PDT)).[1-15] In particular, 6 of the last 8 world record CIGS efficiencies have employed a KF PDT,[1, 3, 4, 12, 16, 17] ultimately advancing the record efficiency from 20.3 to 22.6% in just ~3.5 yr. KF PDT successes in the laboratory have now been extended to commercially-relevant chalcogenized Cu(In,Ga)(Se,S)$_2$ absorbers,[12] full size (0.75 m$^2$) modules,[13] and Cd-free Zn(O,S) buffers.[2, 12, 18] Although the mechanisms responsible for these efficiency improvements are not clear, the KF PDT has been associated with multiple phenomena: *increased* hole concentration (e.g., by consuming In$_{Cu}$ compensating donors to produce K$_{Cu}$ neutral defects[19]),[5-8, 11, 14, 15, 19-23] *decreased* hole concentration (e.g., by consuming Na$_{Cu}$, which produces In$_{Cu}$ compensating donors),[1, 8, 10] Na depletion or formation of soluble Na chemical(s),[1, 5, 7, 8, 10, 13, 14, 22-25] Ga depletion at the surface,[1, 8, 10, 13-15, 24, 26-28] Cu depletion at the surface[7, 13, 15] resulting in better near-surface inversion[1, 8, 10, 29] or decreased valence band energy,[14, 21, 26, 28, 30] grain boundary passivation,[5, 12, 31] general defect passivation,[2, 3, 12, 23] minority carrier lifetime enhancement,[9, 15] morphology changes resulting in increased CdS nucleation sites,[2, 10, 32] general changes in CdS growth,[14, 26, 29] formation of a passivating K-In-Ga-Se[10, 33] or K-In-Se[14, 26-28] interfacial compound, formation of a current blocking interfacial compound,[22] formation of elemental Se at the surface,[14, 27] formation of surface $Cu_{2-x}Se$ and $GaF_3$,[14] consumption of a surface 'ordered vacancy compound,'[14] decreased trap concentration,[24] and modified Cu-Ga-In interdiffusion.[6, 20] Another group co-evaporated KF, In, and Se onto CIGS, forming either $KInSe_2$ or K-doped amorphous $In_2Se_3$, and achieved results similar to a KF PDT.[27] A PDT without KF (just Se) has also been shown to significantly alter CIGS absorbers,[34, 35] calling into question what 'control' absorber is appropriate for KF PDT comparison. High absorber Na and K composition has also been linked to drastically accelerated degradation in photovoltaic (PV) performance,[36] which may undermine the industrial relevance of initial performance gains achieved with high Na and K levels. Specifically, Na and K diffused into and degraded the ZnO layer after 100 h in damp heat and light,[36] underscoring the importance of understanding alkali metal bonding in CIGS.

While the mechanisms underlying KF PDT effects remain uncertain, it has been established that a relatively large amount of K is present at the p-n junction in the most efficient solar cells.[1] These advances have heightened interest in Cu-K-In-Se material with group I-poor (i.e. (K+Cu)/In ~ 0.85) and K-rich (x > 0.30) compositions. In a departure from work utilizing KF precursors and PDTs, Cu, KF, In, and Se were co-evaporated to form $Cu_{1-x}K_xInSe_2$ (CKIS) alloys with K/(K+Cu) composition, or x, varied from 0 to 1.[25, 37, 38] Increasing x in CKIS was found to monotonically decrease the chalcopyrite lattice parameter, increase the band gap, and increase the apparent carrier concentration. Moderate K compositions (0 < x < 0.30) exhibited significantly longer minority carrier lifetimes,[37, 38] relative to x ~ 0 and x ≥ 0.30. Superior PV performance was also observed for x ~ 0.07 for a Ga-alloyed film with Ga/(Ga+In) ~ 0.3.[23] The substrate surface's Na composition was also found to determine the relative amount of CKIS alloy formation during growth, relative to $CuInSe_2$ + $KInSe_2$ mixed-phase formation.[25] Those studies are extended in the present work to include the effect of temperature on phase formation in Cu-K-In-Se with compositions of interest for PV applications. The open questions concerning K bonding in CIGS are thereby addressed through investigating chemistry of the Cu-K-In-Se material system.

Experimental Section

Most CIGS films are grown with Ga/(Ga+In) ~ 0.2 – 0.3 and intentional gradients in cation composition.[1, 39, 40] However, Ga has presently been excluded to simplify data interpretation. A constant rate, single temperature process was also chosen to achieve uniformity in depth and avoid compositional gradients–as KF[6, 20] and Na[41, 42] have been shown to affect cation diffusion in CIGS. Co-evaporation of Cu, KF, In, and Se was performed on substrates of SLG and SLG/Mo ("Mo") at 400°C, 500°C, or 600°C, as previously detailed.[38] The (K+Cu)/In composition was maintained near 0.85 for all films, while K/(K+Cu) was varied between 0 and 1. Deposited film compositions were measured with X-ray fluorescence (XRF), secondary ion mass spectrometry (SIMS), and energy dispersive X-ray spectroscopy (EDS on a transmission electron microscope). As noted,[38] peak overlap between K and In reduced the certainty of XRF and EDS, so film compositions from in situ electron impact emission spectroscopy (EIES) and quartz crystal microbalance data were used unless otherwise noted. Interference between $^{39}K^{41}K^+$ and $^{80}Se^+$ ions was observed in SIMS on films with high K content (x ≥ 0.38), so

$^{76}$Se$^+$ ions were used. Symmetric X-ray diffraction (XRD) was performed with a Rigaku Ultima IV diffractometer to determine structure and assist in phase identification, as previously reported.[38] Standard diffraction patterns were calculated from published phase structures.[38, 43, 44] Film morphology was observed using scanning electron microscopy (SEM). Ultraviolet-visible (UV-visible) spectroscopy was performed to measure transmissivity and reflectivity of SLG/Cu-K-In-Se samples using a Cary 5000 spectrophotometer with a diffuse reflectance integrating sphere. Film thicknesses were measured with a Dektak 8 profilometer. Electron-beam evaporation of ~50 nm Ni, followed by ~3 μm Al through ~1.5 mm$^2$ apertures (normally used for solar cell grids) formed top contacts for current-voltage (IV) measurements. IV measurements were performed on a temperature-controlled stage at 25°C. Room temperature time-resolved photoluminescence (TRPL) was performed on absorber films with a 905 nm laser (1.37 eV) under low-injection conditions, and the response was detected with a near-infrared photomultiplier tube responsive to photons in the range 0.92 to 1.31 eV (details of the fiber optic system have been published elsewhere[45]). Peak TRPL signals were used to estimate majority carrier concentrations by assuming a constant radiative coefficient for all samples, and single exponential functions were fit to decays to estimate minority carrier lifetimes. Films were annealed in vacuum by supplying evaporated Se, and heating at 100°C/min to 600°C, holding for 10 - 80 min, and cooling at 20°C/min. Samples and elemental Se were sealed in a holder with Al windows, which was purged with N$_2$, which was heated up to 600°C while symmetric high temperature XRD (HTXRD) was performed. For ramps, temperature was rapidly increased by 10 or 20°C increments, followed by an 8 min XRD scan at constant temperature. For dwells, the

temperature was rapidly ramped (100°C/min) to the set point, and held for the duration of the scans.

Results

### Effect of Growth Temperature

#### High Na Substrates (SLG)

Baseline films with K/(K+Cu), or $x \sim 0$ ($CuInSe_2$) grown at 500°C had more narrow, intense XRD peaks and more precipitous absorptivity onsets by UV-visible spectroscopy, relative to 400°C growths (not shown)—products of enhanced crystallinity at higher growth temperature. The K-free $CuInSe_2$ films did *not* exhibit shifts in either XRD peaks or extrapolated band gaps with growth temperature. Films of Cu-K-In-Se with K/(K+Cu), or $x \sim 0.38$ were grown on SLG substrates at 500°C and 600°C. The 500°C film was homogenous and had a finer grain structure, possibly indicating an increased nucleation rate, relative to the 600°C film (Figure 1). On the other hand, the 600°C film exhibited segregation of large, planar crystals (right of Figure 1). These precipitates appeared to be $KInSe_2$, based on previous reports.[25, 38] The differences between the 500°C and 600°C growth temperatures on SLG were similar to the differences between SLG and Mo substrates previously reported at 500°C.[25] Films were also grown with $x \sim 0.57$, and room temperature XRD showed that at 500°C the CKIS alloy formed to completion (Figure 2), as inferred from chalcopyrite lattice parameter reduction.[38] The 600°C film had chalcopyrite peaks corresponding to $x \sim 0$ ($CuInSe_2$), as well as $KInSe_2$. The XRD results were corroborated by UV-visible spectroscopy: the 500°C film had a weak absorptivity onset that was nonlinear in the Tauc plot, and yielded a band gap of 1.28 eV (Figure S1), as expected for a CKIS film with x ~ 0.57.[38] The 600°C film had two absorptivity transitions, which were linear in the Tauc plot, and yielded 0.90 and 2.50 eV band gaps—similar to the values for CuInSe$_2$ and KInSe$_2$ (1.05 and 2.71 eV[38]). The nonlinearity in the 500°C film reduced the certainty of its band gap, and the 600°C film's mixed phases violated the assumptions for extrapolating band gap with a Tauc plot, so these values were considered rough.

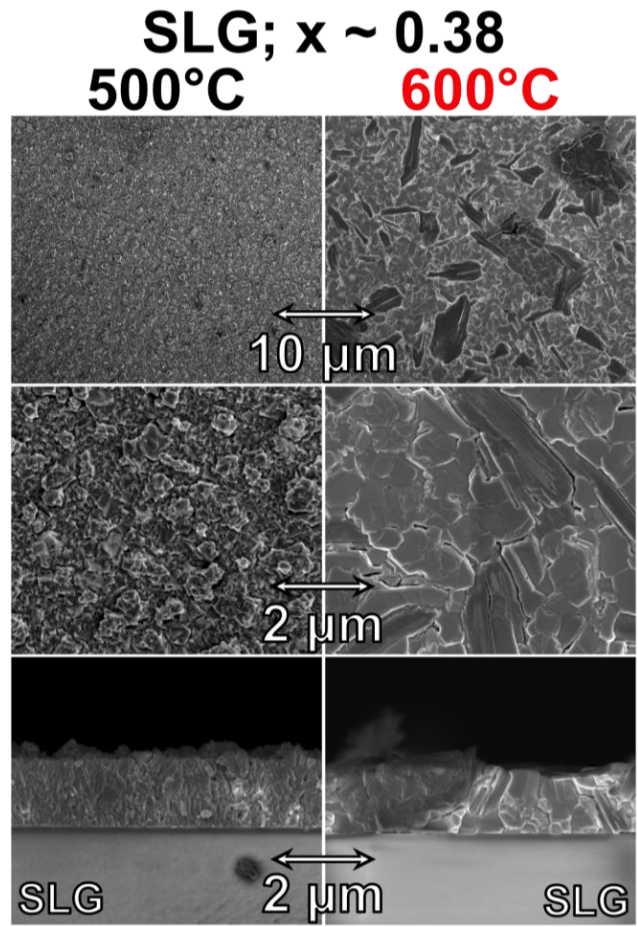

Figure 1. Plan view (top) and cross-sectional (bottom) SEM micrographs of Cu-K-In-Se films with K/(K+Cu), or x ~ 0.38 grown on SLG at 500°C (left) and 600°C (right).

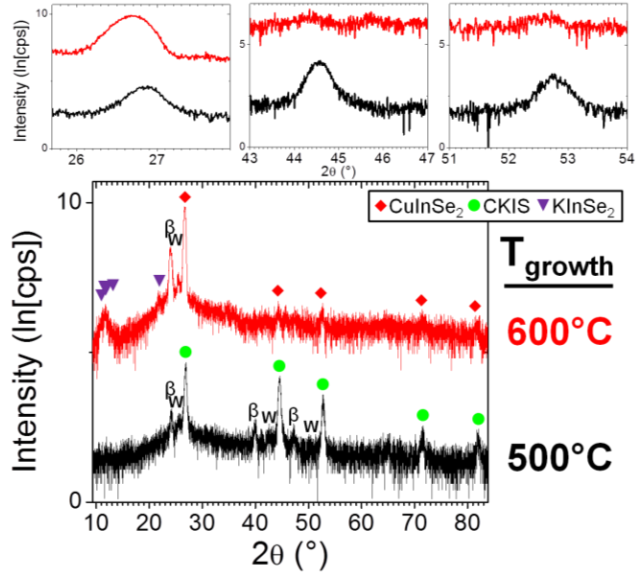

Figure 2. Room temperature symmetric XRD scans from Cu-K-In-Se films with K/(K+Cu), or x ~ 0.38 grown on SLG at 600°C (top; red) and 500°C (bottom; black). CuInSe$_2$, CKIS, and KInSe$_2$ peaks are labeled with red diamonds, green circles, and purple triangles, respectively. 'β' and 'W' peaks are from Cu K$_\beta$ and W impurity in the Cu radiation source.

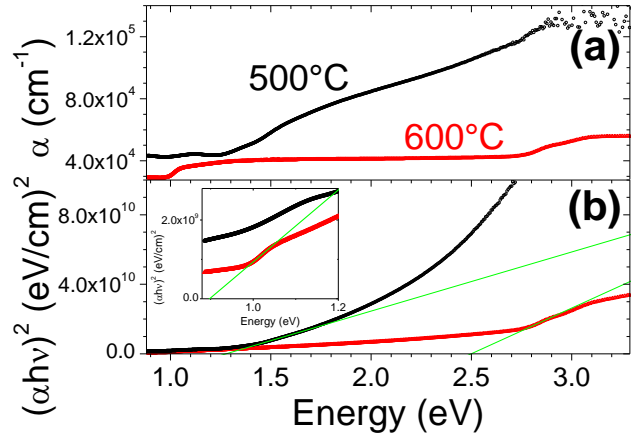

Figure S1. (a) Absorptivity, α, and (b) Tauc plot, $(\alpha h\nu)^2$, versus photon energy for Cu-K-In-Se with K/(K+Cu), or x ~ 0.57 grown on SLG at 500°C (black circles) and 600°C (red triangles). Example least squares fits (green lines in (b) and (b) inset) extrapolated to 1.28, 0.90, and 2.50 eV.

### Low Na Substrates (Mo)

Substrates of Mo were previously shown to establish 3x less Na in the growing Cu-K-In-Se films by SIMS.[25] The effect of temperature on Na-deficient growth was studied for Cu-K-In-Se films with x ~ 0 and 0.38 on Mo at 400°C and 500°C. The x ~ 0 film exhibited enhanced crystallinity at higher growth temperature, but no XRD peak shifts (not shown). The film with x ~ 0.38 in Figure 3 (right) had a composition of 20.5 ± 1.4 Cu, 10.6 ± 1.0 K, 18.1 ± 0.8 In, and 50.9 ± 2.0 Se (at. %) by EDS (x ~ 0.341 ± 0.030), in rough agreement with the values from in situ measurement (13.8 Cu, 8.6 K, 27.6 In, and 50.0 Se (at. %); x ~ 0.376). The film grown at 400°C was homogenous, and had a finer grain structure, relative to 500°C (Figure 3). This could relate to an increased nucleation rate, which would be expected at lower growth temperature. The 500°C film had large, planar $KInSe_2$ precipitates (right of Figure 3). The experiment was repeated with x ~ 0.57, and XRD on the 400°C film showed completely alloyed CKIS, while the 500°C film had mixed-phase $CuInSe_2$ + $KInSe_2$ (Figure 4). Thus, changing growth temperature and substrate Na revealed very similar trends in phase growth.[25] The 400°C film had additional peaks at 17.9 and 29.6° 2θ. These peaks were also observed in a similarly grown x ~ 0.38 film (not shown), only shifted to larger d-spacing. Therefore, they were tentatively assigned to CKIS. Previous reports on bulk $Cu_{0.33}K_{0.67}InSe_2$ crystals found monoclinic symmetry.[46, 47] These XRD peaks could relate to the structural transition from tetragonal $CuInSe_2$ to monoclinic $Cu_{0.33}K_{0.67}InSe_2$, warranting future study.

The resistivity of films grown on Mo at 400°C was measured for Mo/Cu-K-In-Se/Ni stacks with varied K/(K+Cu), or x composition. Increasing x decreased apparent resistivity at 0 V and in reverse bias (Figure S3), and indicated a current blocking barrier for the Mo and/or Ni interface.[38] The resistivities of films grown at 400°C were larger than those grown at 500°C by roughly 2 decades,[38] while the trends with composition changes were very similar for each growth temperature. The correlation between apparent resistivity and x may therefore be unrelated to the intragranular properties of the $CuInSe_2$, CKIS, or $KInSe_2$ crystals. As noted,[38] some undetected grain boundary-segregated phase could dominate the apparent resistivity changes.[48] Bulk crystal resistivity measurements could help discern this effect. Films grown on Mo had superior lifetimes by TRPL, relative to SLG substrates. TRPL lifetimes and carrier concentrations for Cu-K-In-Se films grown on Mo at 400°C, 500°C, and 600°C at different x compositions are in Figure S4. Growth temperature and composition changed the phase constitution of the films and their semiconducting properties, so the mechanisms responsible for the trends in Figure S4 were unclear. Further study with spectrally-resolved TRPL could reveal connections between improved lifetimes and CKIS alloy formation or consumption.

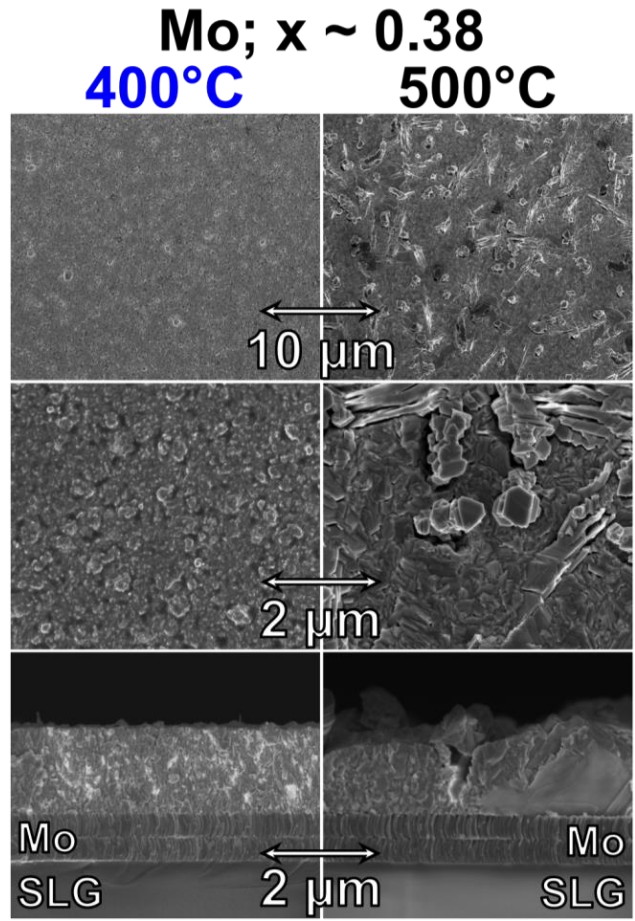

Figure 3. Plan view (top) and cross-sectional (bottom) SEM micrographs of Cu-K-In-Se films with K/(K+Cu), or x ~ 0.38 grown on Mo at 400°C (left) and 500°C (right).

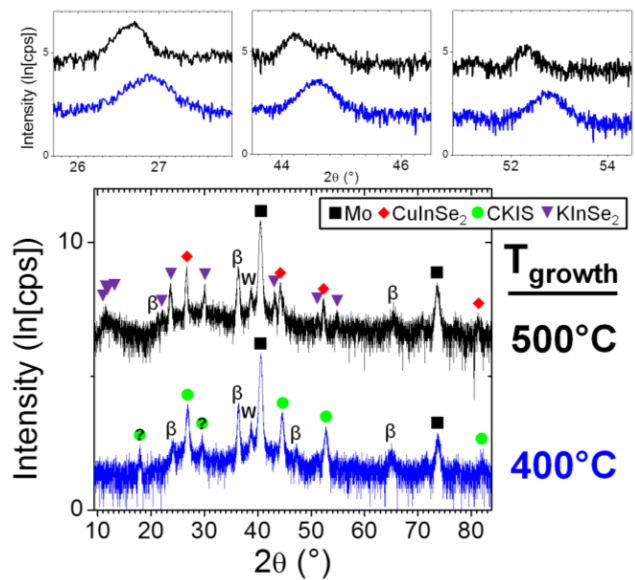

Figure 4. Room temperature symmetric XRD scans from Cu-K-In-Se films with K/(K+Cu), or x ~ 0.57 grown on Mo at 500°C (top; black) and 400°C (bottom; blue). Mo, CuInSe$_2$, CKIS, and KInSe$_2$ peaks are labeled with black squares, red diamonds, green circles, and purple triangles, respectively. A question mark (?) denotes uncertainty in the peak assignment. 'β' and 'W' peaks are from Cu K$_\beta$ and W impurity in the Cu radiation source.

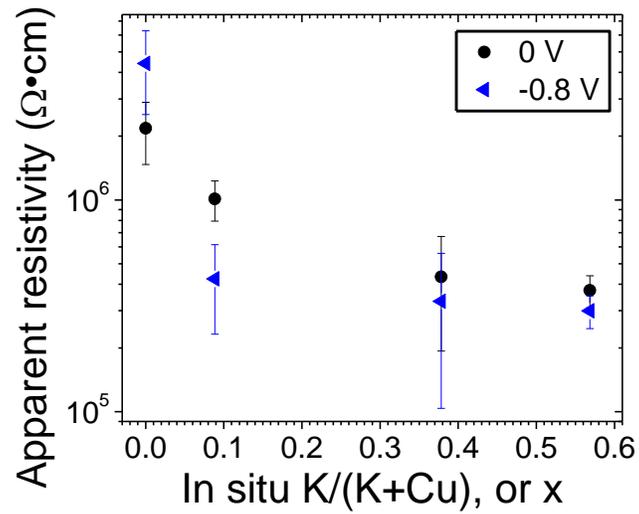

Figure S2. Apparent resistivity at 0 V bias (black circles) and -0.8 V bias (blue triangles) of Mo/CKIS/Ni versus in situ K/(K+Cu), or x composition of the CKIS. CKIS was grown at 400°C, and standard deviations from multiple contacts are shown as error bars.

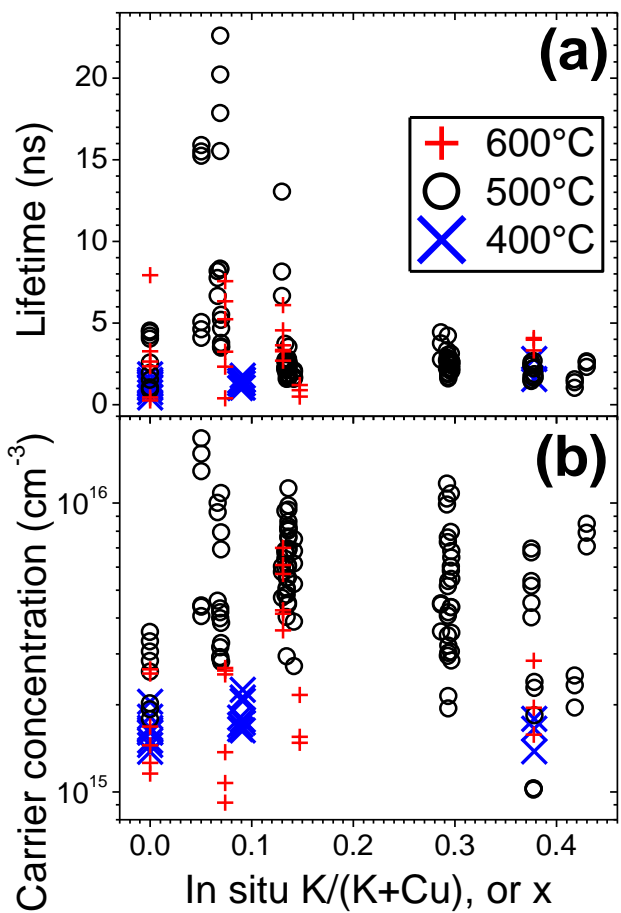

Figure S3. Minority carrier lifetime (a) and majority carrier concentration (b) from TRPL versus in situ K/(K+Cu), or x composition of the Cu-K-In-Se grown on Mo at 600°C (red plusses), 500°C (black circles), and 400°C (blue crosses).

Effect of Annealing

Cu-K-In-Se films were grown on Mo and SLG substrates at 400°C and 500°C, and were then annealed to 600°C for 10 - 80 min under vacuum. Films exhibited increased growth of $KInSe_2$ crystals after annealing (e.g. Figure S4). XRD and UV-visible spectroscopy both showed that CKIS alloys decomposed into $CuInSe_2$ + $KInSe_2$ during anneals (Figures S5 and S6), mirroring behavior at increased growth temperature

(Figures 2, 4, and S1). K-In-Se (x ~ 1) films grown at 500°C predominantly contained KInSe$_2$, although K$_2$In$_{12}$Se$_{19}$ impurities were typical.[38] The band gap of K$_2$In$_{12}$Se$_{19}$ was determined to be 2.25 ± 0.02 eV by UV-visible spectroscopy (not shown), in qualitative agreement with its reported red appearance.[49] Annealing K-In-Se films to 600°C caused some KInSe$_2$ decomposition and K$_2$In$_{12}$Se$_{19}$ formation (Figure S7). Subsequent humid air exposure caused film delamination, preventing extensive characterization. On the other hand, as-grown KInSe$_2$ and K$_2$In$_{12}$Se$_{19}$ films did not decompose after prolonged (30 d) humid air exposure.

HTXRD was used to directly probe the apparent high temperature phase transitions. A CKIS alloy with x ~ 0.57 (grown at 400°C on SLG) was heated to 600°C (Figure 5). KInSe$_2$ peaks appeared in the 290 - 450°C range. Near 410°C, the CKIS developed (220)/(204) texture. At 430°C the background signal increased, possibly due to a phase transition or recrystallization (e.g. the L <-> K$_2$Se + K$_2$Se$_2$ invariant reaction was reported near 430°C[50, 51]). At 420 - 450°C the CKIS peaks abruptly shifted to larger d-spacing, and very intense KInSe$_2$ peaks simultaneously appeared. The KInSe$_2$ peaks at 460°C corresponded with 1% longer a, b, and c lattice parameters. As the temperature ramp continued, however, the KInSe$_2$ lattice continuously compressed isotropically (i.e. peaks at greater 2θ displayed greater shifts). Another peak shifted to smaller 2θ values in Figure 5 (50° 2θ and 520°C), although the source of that shift was unclear.

To further study the peak shifts at high temperature observed in Figure 5, dwell and temperature ramp HTXRD was performed on a K-In-Se film (x ~ 1; grown on SLG at 500°C). The dwell at 550°C (Figure 6 (a)) also exhibited substantial peak shifts, indicative of some kinetic-limited reaction. The temperature ramp showed that the

$K_2In_{12}Se_{19}$ + $KInSe_2$ mixed-phase film recrystallized into $KInSe_2$ near 410 - 450°C (Figure 6 (b)). The high temperature $KInSe_2$ peaks then shifted to smaller d-spacing as temperature and time increased, similar to the shifts of Figures 5 and 6 (a). Other peaks shifted to smaller $2\theta$ values in Figures 6 (a) and (b) (near 31, 42, 50, 55, and 60° $2\theta$ and 0 hr, and 32, 42, 50, 55, and 61° $2\theta$ and 520°C, respectively), although the source of those shifts was unclear. The background increased near 230°C in Figure 6 (b), possibly a result of melting Se. The background increase at 430°C in both Figures 5 and 6 (b) may have emerged from a K-In-Se phase transition or recrystallization.

The XRF composition after anneals showed reduced K content for all x > 0, where more K loss occurred in films with greater initial K content (Figure 7). A $CuInSe_2$ film (x ~ 0) contained more Na and K by SIMS after a 600°C anneal (Table 1)—attributed to increased Na and K out-diffusion from the substrate. However, the CKIS film with x ~ 0.38 showed substantially less K by SIMS after annealing (Na increased similar to the x ~ 0 case). The XRF and SIMS demonstrated that K loss occurred during annealing, although it was kinetically inhibited, as residual K also remained after 80 min anneals. As K loss occurred in addition to other phase transitions, any causal connection was unclear from the data.

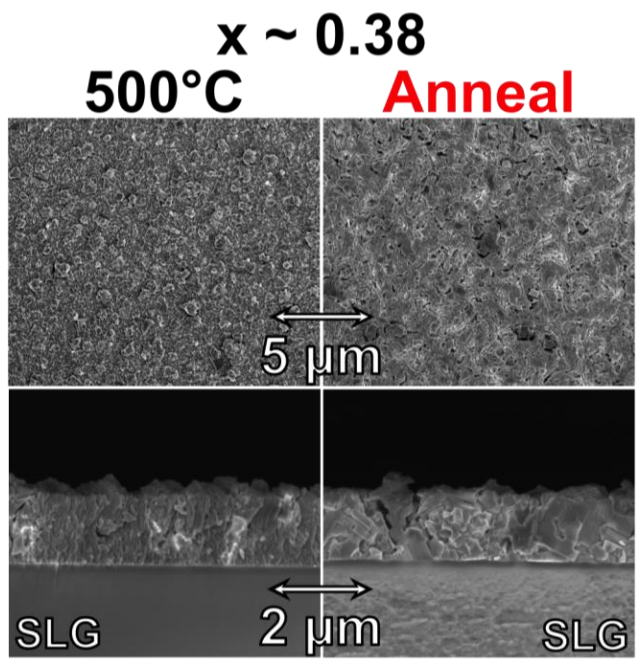

Figure S4. Plan view (top) and cross-sectional (bottom) SEM micrographs of a Cu-K-In-Se film with K/(K+Cu), or x ~ 0.38 grown on SLG at 500°C (left) and then annealed at 600°C for 10 min (right).

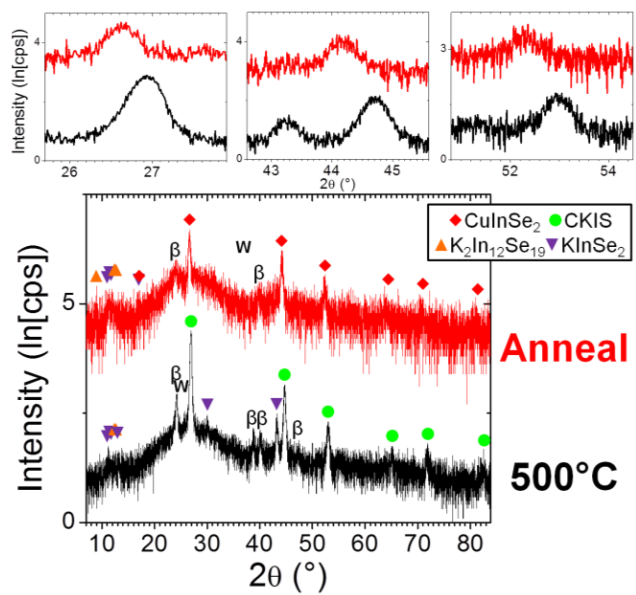

Figure S5. Room temperature symmetric XRD scans from a Cu-K-In-Se film with K/(K+Cu), or x ~ 0.57 grown on SLG at 500°C (bottom; black), and then annealed at

600°C for 10 min (top; red). CuInSe$_2$, CKIS, K$_2$In$_{12}$Se$_{19}$, and KInSe$_2$ peaks are labeled with red diamonds, green circles, orange up triangles, and purple down triangles, respectively. 'β' and 'W' peaks are from Cu K$_β$ and W impurity in the Cu radiation source.

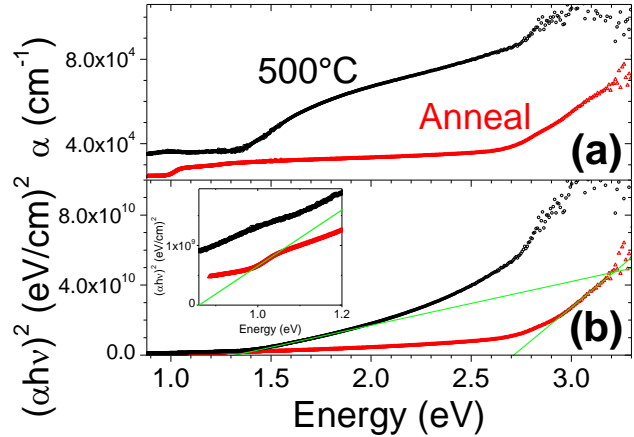

Figure S6. (a) Absorptivity, α, and (b) Tauc plot, (αhν)$^2$, versus photon energy for CKIS with K/(K+Cu), or x ~ 0.57 grown on SLG at 500°C (black circles) and then annealed at 600°C for 10 min (red triangles). Example least squares fits are shown (green lines in (b) and (b) inset) extrapolated to 1.32, 0.86, and 2.71 eV.

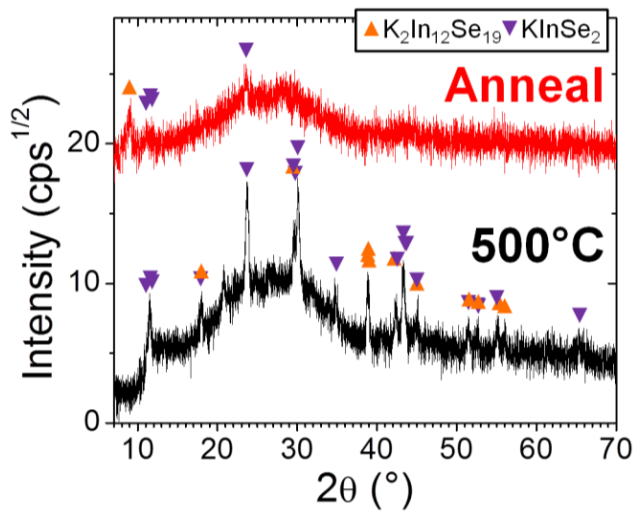

Figure S7. Room temperature symmetric XRD scans from a K-In-Se film (x ~ 1) grown on SLG at 500°C (bottom; black), and then annealed at 600°C for 80 min (top; red).

$K_2In_{12}Se_{19}$ and $KInSe_2$ peaks are labeled with orange up triangles and purple down triangles, respectively.

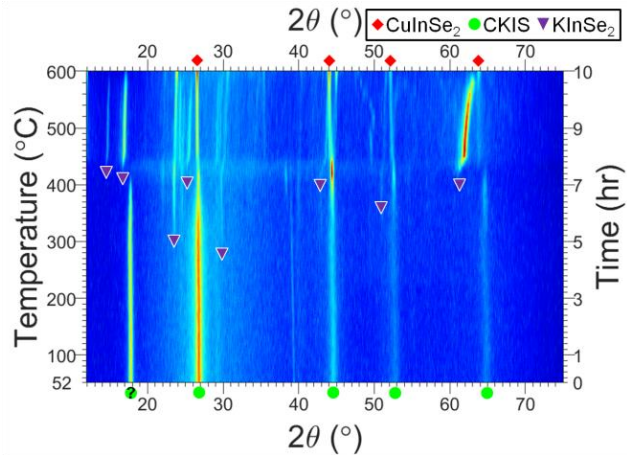

Figure 5. HTXRD temperature ramp for SLG/CKIS with K/(K+Cu), or x ~ 0.57 grown at 400°C. $CuInSe_2$, CKIS, and enlarged-lattice-$KInSe_2$ peaks are labeled with red diamonds, green circles, and purple triangles, respectively. The enlarged-lattice-$KInSe_2$ peaks were calculated with a 1% increase in the a, b, and c lattice parameters. A question mark (?) denotes uncertainty in the peak assignment.

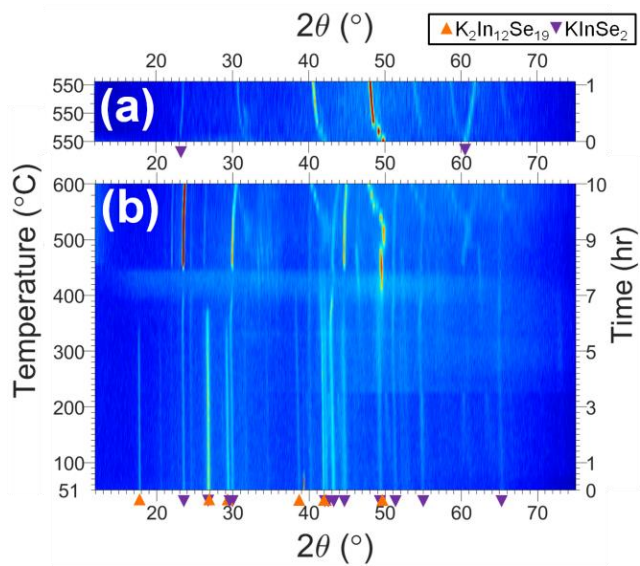

Figure 6. HTXRD dwell at 550°C (a) and temperature ramp (b) for SLG/K-In-Se (i.e. K/(K+Cu), or x ~ 1) grown at 500°C. $K_2In_{12}Se_{19}$ and $KInSe_2$ peaks are labeled with orange up triangles and purple down triangles, respectively.

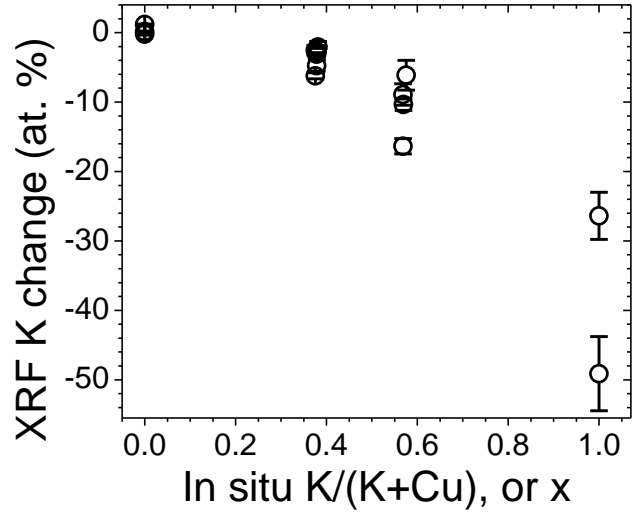

Figure 7. Change in K composition by XRF after annealing Cu-K-In-Se films at 600°C, plotted against in situ K/(K+Cu), or x composition. Films were grown at 400°C and 500°C, on SLG and Mo substrates, and annealed for 10, 20, or 80 min.

Table 1. Na and K SIMS signals integrated over the film and divided by thickness for Cu-K-In-Se films with in situ K/(K+Cu), or x of 0 and 0.38 grown at 500°C on Mo, and then annealed at 600°C for 10 min.

| K/(K+Cu), or x | Sample | Na (cps/μm) | K (cps/μm) |
|---|---|---|---|
| 0 | Annealed | $4.7 \times 10^5$ | $1.4 \times 10^4$ |
| | 500°C growth | $1.6 \times 10^4$ | $4.6 \times 10^3$ |
| 0.38 | Annealed | $2.3 \times 10^6$ | $1.0 \times 10^6$ |
| | 500°C growth | $7.0 \times 10^5$ | $3.0 \times 10^6$ |

Discussion

The effect of growth parameters on the extent of the following net reaction was further solidified:

$$CuInSe_2 + KInSe_2 \longleftrightarrow 2\, Cu_{0.5}K_{0.5}InSe_2 \qquad (1)$$

Switching from low to high Na substrates at 500°C was previously shown to favor the forward reaction. In this study, switching from high to low temperature was shown to further drive the reaction forward. Remarkably similar effects were observed for the following 3 process changes: from Mo to SLG at 500°C, from 600°C to 500°C on SLG, and from 500°C to 400°C on Mo. Further study will be needed to determine the quantitative effects of Na and temperature, as well as the kinetic and/or thermodynamic origins of those effects. The effect of temperature was easily observed on samples with high K/(K+Cu) compositions. At lower K/(K+Cu) compositions, the magnitudes of lattice parameter and band gap shifts would make this effect difficult to observe with the presently employed methods. Future data on lower K/(K+Cu) compositions could help clarify the role of a potential CKIS miscibility gap in the presently observed phase transitions. Upon reversal of reaction (1), the $CuInSe_2$ precipitates could have contained K, where measurements of K solubility in bulk chalcopyrite CIGS crystals would help characterize this.

The SEM, XRD, UV-visible spectroscopy, and HTXRD results on samples annealed under vacuum and atmospheric pressure resembled results for growth temperature, where high temperature annealing reversed reaction (1). Annealing additionally consumed $KInSe_2$ and produced $K_2In_{12}Se_{19}$. However, relative phase amounts of $KInSe_2$ and $K_2In_{12}Se_{19}$ were difficult to distinguish due to peak overlap in XRD (e.g. Figure S5) and similar band gaps: 2.71 eV[37, 52] compared to 2.25 ± 0.02 eV, respectively. It was therefore unclear if the KInSe$_2$ decomposition and K$_2$In$_{12}$Se$_{19}$ formation was also driven by growth temperature, or just annealing. The combined XRF and SIMS results indicated that K loss occurred during the anneals. The reduction in K chemical potential during annealing achieved conditions *unlike* increasing growth temperature, as K chemical potential was held relatively constant in the latter.

Based on the combined XRD, UV-visible spectroscopy, HTXRD, XRF, and SIMS data, the following net reaction was favored at high temperature and reduced K chemical potential:

12 KInSe$_2$ (s) —600°C—> K$_2$In$_{12}$Se$_{19}$ (s) + 5 K$_2$Se (g)   (2)

Here, the gaseous K-containing species was assumed to be K$_2$Se for the sake of illustration, as no K-Se vapor pressure data has been published. The KInSe$_2$ had (53-8)/(51-9)/(356)/(15-8) texture in Figures 5 and 6 (b), while it had (22-2)/(221) texture in Figure 6 (a). However, the contraction of the KInSe$_2$ lattice at high temperature (Figures 5, 6 (a), and 6 (b)) was apparently isotropic, so it may not stem from external stress. On the other hand, an isotropic reduction in KInSe$_2$ lattice volume may be expected for the following reaction:

10 KInSe$_2$ (s) + Se (g) —600°C—> 10 K$_{0.8}$InSe$_2$ (s) + K$_2$Se (g)   (3)

As shown, the formation of vacancies on K lattice sites (Va$_K$) in KInSe$_2$ could have accompanied the evaporative loss of K. In this way, the observed K loss could cause the KInSe$_2$ contraction observed at high temperature, as well as the KInSe$_2$ consumption and K$_2$In$_{12}$Se$_{19}$ formation.

Continuous peak shifts to smaller 2θ (larger d-spacing) were also found at 500-600°C in Figures 5, 6 (a), and 6 (b). These were likely attributable to phase transitions or recrystallization. There were many candidate compounds: $In_4Se_3$, $InSe$, $In_6Se_7$, $In_9Se_{11}$, $In_5Se_7$, 4 polymorphs of $In_2Se_3$,[53] $K_2In_{12}Se_{19}$,[43] the Se analog of $KIn_5S_8$ (if it exists),[54] $KIn_3Se_5$,[52] $KInSe_2$,[44, 46, 52] $K_4In_2Se_5$,[52] $K_{12}In_2Se_9$,[55] $K_9InSe_7$,[55] $K_2Se$, $K_2Se_2$, $K_2Se_4$, $K_2Se_5$, and $K_2Se_6$.[50, 51] The structures of $KIn_3Se_5$, $K_4In_2Se_5$, $K_2Se_2$, $K_2Se_4$, $K_2Se_5$, and $K_2Se_6$ have not been published. Phase diagrams and structural information for the K-In-Se system could help decipher these high temperature peak shifts.

Growth temperature and annealing established different K chemical potentials, but yielded similar results, so the role of substrate Na and temperature may be separate from K chemical potential. Equilibration experiments could assist in determining the specific role of K chemical potential on phase equilibrium in the Cu-K-In-Se system, as well as any thermodynamic component to the extent of reaction (1). While chemical reactions have been shown to occur during KF PDTs,[1, 5, 7, 8, 10, 13-15, 21-30, 33] the specific reactants, products, extents and desirability of the reactions are relatively uncertain. The present work has outlined the stability of $CuInSe_2$ + $KInSe_2$ (relative to CKIS) for various conditions, and contributes to understanding the material chemistry behind the PV performance enhancements/detriments of KF PDTs, guiding more robust process engineering for K introduction, and identifying probable reactants and products in long-term degradation studies.

Conclusions

The effect of temperature on phase formation in the Cu-K-In-Se material system was studied. Increased substrate Na composition was previously shown to drive CKIS alloy formation (relative to $CuInSe_2$ + $KInSe_2$ mixed phases) during co-evaporation.[25] For

Na-rich SLG substrates, changing growth temperature from 600°C to 500°C also drove the formation of CKIS alloys, as evidenced by SEM, XRD, and UV-visible spectroscopy. Na-poor Mo substrates clearly exhibited the same effect on changing growth temperature from 500°C to 400°C. CKIS alloy formation was therefore favored, relative to $CuInSe_2$ + $KInSe_2$, at increased substrate Na and decreased temperature. Films had resistivities, minority carrier lifetimes, and carrier concentrations in line with previous reports.[25, 37, 38] Films grown at 400°C and 500°C were annealed to 600°C, and showed K loss in proportion to initial film K content (XRF and SIMS). Annealing established increased temperature and decreased K chemical potential, while also reversing reaction (1) and advancing reaction (2) forward (by SEM, XRD, UV-visible spectroscopy, and HTXRD). Annealing's effect on reaction (1) was in excellent agreement with increased growth temperature results, despite the differences in K chemical potential for the 2 experiments—evidence that K chemical potential was *not* the mechanism by which temperature and substrate Na affected CKIS alloy formation, although phase diagram studies would be needed to affirm or refute this inference. The $KInSe_2$ lattice exhibited contraction at high temperature that progressed with temperature and time, as well as just time. This could have been a result of $Va_K$ formation, driven by K evaporation (e.g. reaction (3)). However, other unexplained HTXRD peak shifts were observed at high temperature, and more structural and phase equilibrium data for K-In-Se will be needed to better understand these transitions. The present study has identified promising strategies for understanding and ultimately engineering K bonding in CIGS to maximize initial and long-term PV performance: Process conditions were used to establish relative $CuInSe_2$, CKIS, $KInSe_2$, and $K_2In_{12}Se_{19}$ phase amounts. The growth trends may prove useful for predicting the presence or absence of minor phase impurities below typical detection limits—impurities that could dominate PV-relevant electronic properties. Furthermore, the results lay a foundation for identifying the thermodynamic and/or kinetic mechanisms ultimately controlling the formation of each Cu-K-In-Se phase, as well as predicting probable degradation routes in reliability studies.

Acknowledgement

The authors thank Stephen Glynn for a key suggestion, Kannan Ramanathan for leadership, Lorelle Mansfield, Carolyn Beall, Karen Bowers, and Stephen Glynn for assistance with experiments, Clay DeHart for contact deposition, Matt Young for SIMS, and Bobby To for SEM. TEM/EDS and HTXRD were performed at the Center for Nanophase Materials Sciences at Oak Ridge National Laboratory. The work was supported by the U.S. Department of Energy under contract DE-AC36-08GO28308.

Supporting Information Description

Absorptivity data for CKIS with x ~ 0.57 grown at 500°C and 600°C, resistivity data for Mo/CKIS/Ni grown at 400°C at varied x composition, TRPL data for CKIS grown at 400°C, 500°C, and 600°C as a function of x composition, SEM micrographs of CKIS with x ~ 0.38 before and after annealing, XRD of CKIS with x ~ 0.57 before and after annealing, absorptivity data of CKIS with x ~ 0.57 before and after annealing, and XRD of CKIS with x ~ 1 before and after annealing are shown.

References

1. Chirilă, A.; Reinhard, P.; Pianezzi, F.; Bloesch, P.; Uhl, A. R.; Fella, C.; Kranz, L.; Keller, D.; Gretener, C.; Hagendorfer, H.; Jaeger, D.; Erni, R.; Nishiwaki, S.; Buecheler, S.; Tiwari, A. N., Potassium-induced surface modification of Cu(In,Ga)Se$_2$ thin films for high-efficiency solar cells. *Nat Mater* **2013,** 12, (12), 1107-1111.
2. Friedlmeier, T. M.; Jackson, P.; Bauer, A.; Hariskos, D.; Kiowski, O.; Wuerz, R.; Powalla, M., Improved Photocurrent in Cu(In,Ga)Se$_2$ Solar Cells: From 20.8% to 21.7% Efficiency with CdS Buffer and 21.0% Cd-Free. *Photovoltaics, IEEE Journal of* **2015,** 5, (5), 1487-1491.
3. Jackson, P.; Hariskos, D.; Wuerz, R.; Kiowski, O.; Bauer, A.; Friedlmeier, T. M.; Powalla, M., Properties of Cu(In,Ga)Se$_2$ solar cells with new record efficiencies up to 21.7%. *physica status solidi (RRL) – Rapid Research Letters* **2015,** 9, (1), 28-31.
4. Jackson, P.; Hariskos, D.; Wuerz, R.; Wischmann, W.; Powalla, M., Compositional investigation of potassium doped Cu(In,Ga)Se$_2$ solar cells with efficiencies up to 20.8%. *physica status solidi (RRL) – Rapid Research Letters* **2014,** 8, (3), 219-222.
5. Laemmle, A.; Wuerz, R.; Powalla, M., Efficiency enhancement of Cu(In,Ga)Se$_2$ thin-film solar cells by a post-deposition treatment with potassium fluoride. *physica status solidi (RRL) – Rapid Research Letters* **2013,** 7, (9), 631-634.
6. Laemmle, A.; Wuerz, R.; Powalla, M., Investigation of the effect of potassium on Cu(In,Ga)Se$_2$ layers and solar cells. *Thin Solid Films* **2015,** 582, 27-30.
7. Mansfield, L. M.; Noufi, R.; Muzzillo, C. P.; DeHart, C.; Bowers, K.; To, B.; Pankow, J. W.; Reedy, R. C.; Ramanathan, K., Enhanced performance in Cu(In,Ga)Se$_2$ solar cells fabricated by the two-step selenization process with a potassium fluoride postdeposition treatment. *Photovoltaics, IEEE Journal of* **2014,** 4, (6), 1650-1654.
8. Pianezzi, F.; Reinhard, P.; Chirilă, A.; Bissig, B.; Nishiwaki, S.; Buecheler, S.; Tiwari, A. N., Unveiling the effects of post-deposition treatment with different alkaline elements on the electronic properties of CIGS thin film solar cells. *Physical Chemistry Chemical Physics* **2014,** 16, (19), 8843-8851.
9. Pianezzi, F.; Reinhard, P.; Chirilă, A.; Nishiwaki, S.; Bissig, B.; Buecheler, S.; Tiwari, A. N., Defect formation in Cu(In,Ga)Se$_2$ thin films due to the presence of potassium during growth by low temperature co-evaporation process. *Journal of Applied Physics* **2013,** 114, (19), 194508-8.
10. Reinhard, P.; Bissig, B.; Pianezzi, F.; Avancini, E.; Hagendorfer, H.; Keller, D.; Fuchs, P.; Döbeli, M.; Vigo, C.; Crivelli, P.; Nishiwaki, S.; Buecheler, S.; Tiwari, A. N., Features of KF and NaF Postdeposition Treatments of Cu(In,Ga)Se$_2$ Absorbers for High Efficiency Thin Film Solar Cells. *Chemistry of Materials* **2015,** 27, (16), 5755-5764.
11. Raguse, J. M.; Muzzillo, C. P.; Sites, J. R., Effects of sodium and potassium on the photovoltaic performance of CIGS solar cells. *Journal of Photovoltaics* **2016**, submitted.
12. Kamada, R.; Yagioka, T.; Adachi, S.; Handa, A.; Tai, K. F.; Kato, T.; Sugimoto, H. In *New world record Cu(In,Ga)(Se,S)$_2$ thin film solar cell efficiency beyond 22%*, Photovoltaic Specialist Conference (PVSC), 2016 IEEE 43rd, Portland, OR, 2016; Portland, OR, 2016; p in press.

13. Lundberg, O.; Wallin, E.; Gusak, V.; Södergren, S.; Chen, S.; Lotfi, S.; Chalvet, F.; Malm, U.; Kaihovirta, N.; Mende, P.; Jaschke, G.; Kratzert, P.; Joel, J.; Skupinski, M.; Lindberg, P.; Jarmar, T.; Lundberg, J.; Mathiasson, J.; Stolt, L. In *Improved CIGS Modules by KF Post Deposition Treatment and Reduced Cell-to-Module Losses*, 43rd IEEE Photovoltaic Specialists Conference, Portland, OR, June 5-10, 2016, 2016; Portland, OR, 2016; pp 1-4.
14. Lepetit, T. Influence of KF post deposition treatment on the polycrystalline Cu(In,Ga)Se$_2$/CdS heterojunction formation for photovoltaic application. Université de Nantes, Nantes, FR, 2015.
15. Khatri, I.; Fukai, H.; Yamaguchi, H.; Sugiyama, M.; Nakada, T., Effect of potassium fluoride post-deposition treatment on Cu(In,Ga)Se$_2$ thin films and solar cells fabricated onto sodalime glass substrates. *Solar Energy Materials and Solar Cells* **2016,** 155, 280-287.
16. NREL Best Research-Cell Efficiencies. [http://www.nrel.gov/ncpv/images/efficiency_chart.jpg](http://www.nrel.gov/ncpv/images/efficiency_chart.jpg) (June 17, 2016),
17. Press-release, ZSW Sets New World Record for Thin-Film Solar Cells. In Zentrum für Sonnenenergie- und Wasserstoff-Forschung Baden-Württemberg: Stuttgart, DE, 2016.
18. Hariskos, D.; Jackson, P.; Hempel, W.; Paetel, S.; Spiering, S.; Menner, R.; Wischmann, W.; Powalla, M. In *Method for a high-rate solution deposition of Zn(O,S) buffer layer for high efficiency Cu(In,Ga)Se$_2$-based solar cells*, 43rd IEEE Photovoltaic Specialists Conference, Portland, OR, June 5-10, 2016, 2016; Portland, OR, 2016; pp 1-6.
19. Contreras, M. A.; Egaas, B.; Dippo, P.; Webb, J.; Granata, J.; Ramanathan, K.; Asher, S.; Swartzlander, A.; Noufi, R. In *On the role of Na and modifications to Cu(In,Ga)Se$_2$ absorber materials using thin-MF (M=Na, K, Cs) precursor layers*, Photovoltaic Specialists Conference, 1997., Conference Record of the Twenty-Sixth IEEE, 29 Sep-3 Oct 1997, 1997; 1997; pp 359-362.
20. Wuerz, R.; Eicke, A.; Kessler, F.; Paetel, S.; Efimenko, S.; Schlegel, C., CIGS thin-film solar cells and modules on enamelled steel substrates. *Solar Energy Materials and Solar Cells* **2012,** 100, 132-137.
21. Jeong, G. S.; Cha, E. S.; Moon, S. H.; Ahn, B. T., Effect of KF Treatment of Cu(In,Ga)Se$_2$ Thin Films on the Photovoltaic Properties of CIGS Solar Cells. *Current Photovoltaic Research* **2015,** 3, (2), 65-70.
22. Son, Y.-S.; Kim, W. M.; Park, J.-K.; Jeong, J.-h., KF Post Deposition Treatment Process of Cu(In,Ga)Se$_2$ Thin Film Effect of the Na Element Present in the Solar Cell Performance. *Current Photovoltaic Research* **2015,** 3, (4), 130-134.
23. Raguse, J. M.; Muzzillo, C. P.; Sites, J. R.; Mansfield, L., Effects of Sodium and Potassium on the Photovoltaic Performance of CIGS Solar Cells. *Journal of Photovoltaics* **2016**, in press.
24. Karki, S.; Paul, P.; Rajan, G.; Ashrafee, T.; Aryal, K.; Pradhan, P.; Collins, R. W.; Rockett, A. A.; Grassman, T. J.; Ringel, S. A.; Arehart, A. R.; Marsillac, S. In *In-situ and Ex-situ Characterizations of CIGS Solar Cells with KF Post Deposition Treatment*, 43rd IEEE Photovoltaic Specialists Conference, Portland, OR, June 5-10, 2016, 2016; Portland, OR, 2016; pp 1-6.
25. Muzzillo, C. P.; Tong, H. M.; Anderson, T. J., The Effect of Na on Cu-K-In-Se Thin Film Growth. *Journal of Crystal Growth* **2016**, in press.

26. Handick, E.; Reinhard, P.; Wilks, R. G.; Pianezzi, F.; Félix, R.; Gorgoi, M.; Kunze, T.; Buecheler, S.; Tiwari, A. N.; Bär, M. In *NaF/KF Post-Deposition Treatments and their Influence on the Structure of Cu(In,Ga)Se$_2$ Absorber Surfaces*, 43rd IEEE Photovoltaic Specialists Conference, Portland, OR, June 5-10, 2016, 2016; Portland, OR, 2016; pp 1-5.
27. Lepetit, T.; Harel, S.; Arzel, L.; Ouvrard, G.; Barreau, N. In *Co-evaporated KInSe$_2$: a fast alternative to KF post-deposition treatment in high efficiency Cu(In,Ga)Se$_2$ thin film solar cells*, 43rd IEEE Photovoltaic Specialists Conference, Portland, OR, June 5-10, 2016, 2016; Portland, OR, 2016; pp 1-4.
28. Handick, E.; Reinhard, P.; Alsmeier, J.-H.; Köhler, L.; Pianezzi, F.; Krause, S.; Gorgoi, M.; Ikenaga, E.; Koch, N.; Wilks, R. G.; Buecheler, S.; Tiwari, A. N.; Baer, M., Potassium post-deposition treatment-induced band gap widening at Cu(In,Ga)Se$_2$ surfaces – Reason for performance leap? *ACS Applied Materials & Interfaces* **2015,** 7, (49), 27414-27420.
29. Umsur, B.; Calvet, W.; Steigert, A.; Lauermann, I.; Gorgoi, M.; Prietzel, K.; Greiner, D.; Kaufmann, C. A.; Unold, T.; Lux-Steiner, M., Investigation of the potassium fluoride post deposition treatment on the CIGSe/CdS interface using hard x-ray photoemission spectroscopy - a comparative study. *Physical Chemistry Chemical Physics* **2016**.
30. Pistor, P.; Greiner, D.; Kaufmann, C. A.; Brunken, S.; Gorgoi, M.; Steigert, A.; Calvet, W.; Lauermann, I.; Klenk, R.; Unold, T.; Lux-Steiner, M.-C., Experimental indication for band gap widening of chalcopyrite solar cell absorbers after potassium fluoride treatment. *Applied Physics Letters* **2014,** 105, (6), 063901-4.
31. Maeda, T.; Kawabata, A.; Wada, T., First-principles study on alkali-metal effect of Li, Na, and K in CuInSe$_2$ and CuGaSe$_2$. *Japanese Journal of Applied Physics* **2015,** 54, (8S1), 08KC20.
32. Friedlmeier, T. M.; Jackson, P.; Kreikemeyer-Lorenzo, D.; Hauschild, D.; Kiowski, O.; Hariskos, D.; Weinhardt, L.; Heske, C.; Powalla, M. In *A Closer Look at Initial CdS Growth on High-Efficiency Cu(In,Ga)Se$_2$ Absorbers Using Surface-Sensitive Methods*, 43rd IEEE Photovoltaic Specialists Conference, Portland, OR, June 5-10, 2016, 2016; Portland, OR, 2016; pp 1-5.
33. Reinhard, P.; Bissig, B.; Pianezzi, F.; Hagendorfer, H.; Sozzi, G.; Menozzi, R.; Gretener, C.; Nishiwaki, S.; Buecheler, S.; Tiwari, A. N., Alkali-Templated Surface Nanopatterning of Chalcogenide Thin Films: A Novel Approach Toward Solar Cells with Enhanced Efficiency. *Nano Letters* **2015,** 15, (5), 3334-3340.
34. Barreau, N.; Zabierowski, P.; Arzel, L.; Igalson, M.; Macielak, K.; Urbaniak, A.; Lepetit, T.; Painchaud, T.; Dönmez, A.; Kessler, J., Influence of post-deposition selenium supply on Cu(In,Ga)Se$_2$-based solar cell properties. *Thin Solid Films* **2015,** 582, 43-46.
35. Shin, Y. M.; Lee, C. S.; Shin, D. H.; Kwon, H. S.; Park, B. G.; Ahn, B. T., Surface modification of CIGS film by annealing and its effect on the band structure and photovoltaic properties of CIGS solar cells. *Current Applied Physics* **2015,** 15, (1), 18-24.
36. Theelen, M.; Hans, V.; Barreau, N.; Steijvers, H.; Vroon, Z.; Zeman, M., The impact of alkali elements on the degradation of CIGS solar cells. *Progress in Photovoltaics: Research and Applications* **2015,** 23, (5), 537-545.
37. Muzzillo, C. P. Chalcopyrites for Solar Cells: Chemical Vapor Deposition, Selenization, and Alloying. Dissertation, University of Florida, Gainesville, FL, 2015.

38. Muzzillo, C. P.; Mansfield, L. M.; Ramanathan, K.; Anderson, T. J., Properties of Cu$_{1-x}$K$_x$InSe$_2$ alloys. *Journal of Materials Science* **2016,** 51, (14), 6812-6823.
39. Muzzillo, C. P.; Mansfield, L. M.; Dehart, C.; Bowers, K.; Reedy, R. C.; To, B.; Noufi, R.; Ramanathan, K.; Anderson, T. J. In *The effect of Ga content on the selenization of co-evaporated CuGa/In films and their photovoltaic performance*, Photovoltaic Specialist Conference (PVSC), 2014 IEEE 40th, Denver, CO, 8-13 June 2014, 2014; Denver, CO, 2014; pp 1649-1654.
40. Muzzillo, C. P.; Mansfield, L. M.; DeHart, C.; Bowers, K.; Reedy, R. C.; To, B.; Ramanathan, K.; Anderson, T. J. In *Differences between CuGa/In and Cu/Ga/In films for selenization*, Photovoltaic Specialist Conference (PVSC), 2015 IEEE 42nd, New Orleans, LA, 14-19 June 2015, 2015; New Orleans, LA, 2015; pp 1-6.
41. Ishizuka, S.; Yamada, A.; Islam, M. M.; Shibata, H.; Fons, P.; Sakurai, T.; Akimoto, K.; Niki, S., Na-induced variations in the structural, optical, and electrical properties of Cu(In,Ga)Se$_2$ thin films. *Journal of Applied Physics* **2009,** 106, (3), 034908-6.
42. Lundberg, O.; Lu, J.; Rockett, A.; Edoff, M.; Stolt, L., Diffusion of indium and gallium in Cu(In,Ga)Se$_2$ thin film solar cells. *Journal of Physics and Chemistry of Solids* **2003,** 64, (9–10), 1499-1504.
43. Schlosser, M.; Reiner, C.; Deiseroth, H.-J.; Kienle, L., K$_2$In$_{12}$Se$_{19}$, a Complex New Structure Type Based on Icosahedral Units of Se$^{2-}$. *European Journal of Inorganic Chemistry* **2001,** 2001, (9), 2241-2247.
44. Wang, P.; Huang, X.-Y.; Liu, Y.-L.; Wei, Y.-G.; Li, J.; Guo, H.-Y., Solid state synthesis at intermediate temperature and structural characterization of KInSe$_2$. *Acta Chimica Sinica* **2000,** 58, (8), 1005-1008.
45. Repins, I. L.; Egaas, B.; Mansfield, L. M.; Contreras, M. A.; Muzzillo, C. P.; Beall, C.; Glynn, S.; Carapella, J.; Kuciauskas, D., Fiber-fed time-resolved photoluminescence for reduced process feedback time on thin-film photovoltaics. *Review of Scientific Instruments* **2015,** 86, 013907-7.
46. Ma, H.-W.; Guo; Wang, M.-S.; GuoZhou; Lin, S.-H.; Dong, Z.-C.; Huang, J.-S., K$_2$MM'$_3$Se$_6$ (M = Cu, Ag; M' = Ga, In), A New Series of Metal Chalcogenides with Chain−Sublayer−Chain Slabs: $_\infty^1$[M'Se$_4$]−$_\infty^2$[(MSe$_4$)(M'Se$_4$)]−$_\infty^1$[M'Se$_4$]. *Inorganic Chemistry* **2003,** 42, (4), 1366-1370.
47. Wang, G.-H.; Guo, H.-Y., Solid-state synthesis of K$_2$CuIn$_3$Se$_6$ and its crystal structure and characterization. *Journal of Beijing University of Chemical Technology* **2004,** 31, (2), 73-77.
48. Orton, J. W.; Powell, M. J., The Hall effect in polycrystalline and powdered semiconductors. *Reports on Progress in Physics* **1980,** 43, (11), 1263-1307.
49. Kienle, L.; Simon, A., Microdomains and Diffuse Scattering in K$_2$In$_{12}$Se$_{19}$. *Journal of Solid State Chemistry* **2001,** 161, (2), 385-395.
50. Klemm, W.; Sodomann, H.; Langmesser, P., Beiträge Zur Kenntnis der Alkalimetallchalkogenide. *Zeitschrift für anorganische und allgemeine Chemie* **1939,** 241, (2-3), 281-304.
51. Sangster, J.; Pelton, A. D., The K-Se (Potassium-Selenium) System. *Journal of Phase Equilibria* **1997,** 18, (2), 177-180.
52. Kish, Z. Z.; Lazarev, V. B.; Peresh, E. Y.; Semrad, E. E., Compounds in In$_2$Se$_3$-K$_2$Se. *Neorganicheskie Materialy* **1988,** 24, (10), 1602-1605.

53. Li, J.-B.; Record, M.-C.; Tedenac, J.-C., A thermodynamic assessment of the In-Se system. *Zeitschrift für Metallkunde* **2003,** 94, (4), 381-389.
54. Deiseroth, H. J., Splitpositionen für Alkalimetallkationen in den Thioindaten MIn$_5$S$_8$ (M = K, Rb, Cs)? *Zeitschrift für Kristallographie - Crystalline Materials* **1986,** 177, (1-4), 307-314.
55. Heine, J.; Dehnen, S., Synthesis and Characterization of the First Salts of the ortho-Selenidoindate Anion [InSe$_4$]$^{5-}$. *Zeitschrift für anorganische und allgemeine Chemie* **2008,** 634, (12-13), 2303-2308.